\documentclass[aps,prc,twocolumn,showpacs,preprintnumbers,amsmath,amssymb
,nofootinbib]{revtex4-1}
\usepackage{graphicx}
\usepackage{bm}
\usepackage{booktabs}
\usepackage{subfigure}
\usepackage[usenames]{color}

\def\lsim{\buildrel < \over {_{\sim}}}
\def\gsim{\buildrel > \over {_{\sim}}}

\newcommand{\beq}{\begin{equation}}
\newcommand{\eeq}{\end{equation}}
\newcommand{\be}{\begin{eqnarray}}
\newcommand{\ee}{\end{eqnarray}}

\usepackage{sublabel}
\usepackage{epsf}
\usepackage{wrapfig}
\usepackage{epsfig}
\usepackage[usenames]{color}
\usepackage{pstricks}
\usepackage{amssymb, graphics}
\usepackage{indentfirst}
\usepackage{fancyhdr}
\usepackage[latin1]{inputenc}
\usepackage[english]{babel}
\usepackage{amssymb}  
\usepackage{amsmath}  
\usepackage{latexsym} 
\usepackage{amsthm}   
\usepackage{subfigure}
\usepackage{color}
\usepackage{natbib,hyperref}

\begin{document}
\title{Neutral current interactions of low-energy neutrinos in dense neutron matter}
\author{Alessandro Lovato$^{1,2}$}
\author{Omar Benhar$^{3}$}
\altaffiliation[on leave from ]{INFN and Department of Physics, ``Sapienza'' Universit\`a di Roma, I-00185 Roma, Italy.}
\author{Stefano Gandolfi$^{4}$}
\author{Cristina Losa$^{5}$}

\affiliation
{
$^1$ Argonne Leadership Computing Facility, Argonne National Laboratory, Argonne, IL 60439, USA\\
$^2$ Physics Division, Argonne National Laboratory, Argonne, IL-60439, USA \\
$^3$ Center for Neutrino Physics, Virginia Polytechnic Institute and State University, 
Blacksburg, VA 24061, USA  \\
$^4$ Theoretical Division, Los Alamos National Laboratory, Los Alamos, NM 87545, USA \\
$^5$ International School for Advanced Studies (SISSA), I-34136 Trieste, Italy 
}
\date{\today}
\begin{abstract}
We report the results of a calculation of the response of cold neutron matter to neutral-current interactions with low energy neutrinos, 
carried out using an effective interaction and effective operators 
consistently derived within the formalism of Correlated Basis Functions. The neutrino mean free path obtained from the calculated responses turns out to be 
strongly affected by both short and long range correlations, leading to a sizable increase with respect to the prediction of the Fermi gas model.
The consistency between the proposed approach and Landau theory of normal Fermi liquids also has been investigated, using a set of Landau parameters
obtained from the matrix elements of the effective interaction.  
\end{abstract}
\maketitle

\section {Introduction}
\label{intro}

Over the past decade, {\em ab initio} nuclear many-body approaches -- based on dynamical models strongly
constrained by the properties  of {\em exactly solvable} two- and three-nucleon systems -- have reached
the degree of maturity required to provide a consistent description of a variety of nuclear matter properties 
other than the zero-temperature equation of state, the knowledge of which is needed for the description 
of neutron star structure and dynamics.  

In this context, an important role was played by the development of effective interactions derived from realistic 
nuclear Hamiltonians using the formalism of Correlated Basis Functions (CBF) 
and cluster expansion techniques \cite{shannon_veff,BV,Lovato13}. 
Use of these effective interactions allows one to evaluate the nucleon-nucleon scattering rate in nuclear matter, needed to obtain the transport 
coefficients from Boltzmann's equation \cite{BV,Gmat}, as well as to treat short- and long-range correlation effects in the 
nuclear response on the same footing \cite{shannon_veff,Benhar_Farina,Lovato13}. 

The response of neutron star matter to weak interactions determines its opacity to neutrinos, which in turn drives the energy loss 
caused by the flux of neutrinos leaving the star. As a consequence, opacity, conveniently parametrized in terms of the 
neutrino mean free path, is a key element for the description of neutron star cooling. The neutrino mean free path is in fact one 
of the critical inputs required for large-scale simulations of neutrino transport.

In Ref. \cite{shannon_veff} the neutrino mean free path in cold isospin symmetric nuclear matter has been obtained from the 
weak response computed using a CBF effective
interaction, derived taking into account the contribution of two-nucleon clusters and using a nuclear Hamiltonian including two-nucleon 
interactions only. The authors of Ref. \cite{BV} improved on the model of Ref. \cite{shannon_veff} by adding the effects of 
three-nucleon interaction, which are known to become dominant at densities larger than the nuclear saturation density, described through 
a density-dependent modification of the nucleon-nucleon (NN) potential. 
The resulting effective interaction has been used in Ref. \cite{Benhar_Farina} to carry out a detailed analysis of the effects of short- and long-range 
correlations on the response of isospin symmetric nuclear matter to charged current weak interactions. 

In Ref. \cite{Lovato13}, the CBF effective interaction has been further improved taking into account the contributions of three-nucleon 
clusters, the inclusion of which allows one to adopt a fully microscopic model of 
three-nucleon forces.
The resulting effective interaction and the corresponding effective operators have been used to carry out a calculation 
of the density and spin-density responses of cold isospin symmetric nuclear matter.

The matrix elements of the effective interaction of Ref.~\cite{BV} have also been used to obtain the set of Landau parameters, 
which was in turn employed to carry out the calculation of the neutron matter responses within the framework of a conceptually different 
scheme  \cite{BCL, andrea_thesis}. 

In this article, we apply the approach developed in Ref.~\cite{Lovato13} to cold neutron 
matter, and use the resulting response functions to obtain the neutrino 
mean free path. We also compare the responses computed using the CBF effective
operators and effective interaction to the ones obtained from the set of 
Landau parameters corresponding to the effective interaction of Ref. \cite{Lovato13}.
Finally, in order to gauge the validity of the approximations involved in our
calculations, we compare the sum rules resulting from energy integration of the  
responses to those obtained evaluating the 
ground state expectation values of the density and spin-density fluctuation operators.

The main elements of the formalism employed to obtain the response functions and the 
CBF effective interaction and effective operators are reviewed in Section \ref{formalism}, while  
Section \ref{totresults} is devoted to the discussion of the results of numerical calculations, including the 
density and spin-density responses, the neutrino mean free path and the static structure functions. 
Finally,  in Section \ref{conclusions} we state the conclusions and outline the prospects of our work.






\section{Formalism}
\label{formalism}

\subsection{Neutral current interactions}
\label{NC}

In the low energy limit of Weinberg-Salam's theory of neutral current weak interactions, the rate of the process
in which a neutrino traveling through neutron matter with four-momentum $k\equiv(E,\mathbf{k})$ is scattered to a state of four-momentum   
$k^\prime=(E^\prime,\mathbf{k}^\prime)$ can be written in the form
\begin{equation}
W(\mathbf{q},\omega)= \frac{G_{F}^2}{4 \pi^2} \frac{1}{ E E^\prime} L_{\mu\nu}W^{\mu\nu}\ ,
\label{eq:scattering_rate}
\end{equation}
where $G_F$ is the Fermi coupling constant and $q=k-k^\prime \equiv (\omega,\mathbf{q})$ denotes the four-momentum transfer.

Neglecting the neutrino mass, the lepton tensor $L_{\mu\nu}$ can be expressed in terms of the lepton kinematical variables according to
\begin{equation}
L_{\mu\nu}=k_\mu k^{\prime}_\nu+k_\nu k^{\prime}_\mu - g_{\mu\nu} (k k^\prime) + i \epsilon_{\mu\alpha\nu\beta} \ k^\alpha k^{\prime\,\beta} \ ,
\end{equation}
where $g_{\mu\nu} = {\rm diag}({1,-1,-1,-1})$ and $\epsilon_{\mu\alpha\nu\beta}$ is the fully antisymmetric Levi-Civita tensor. 

All information on strong interaction physics is contained in the hadronic tensor  
\begin{equation}
\label{def:W}
W^{\mu\nu}=\sum_n \langle  \Psi_0| {J_{Z}^\mu}^\dagger | \Psi_n \rangle  \langle  \Psi_n| J_{Z}^\nu |   \Psi_0 \rangle \delta(\omega+E_0-E_n) \ ,
\end{equation}
the definition of which involves the neutron matter initial and final states, as well as its current operator.

In the non--relativistic limit, corresponding to $|\mathbf{q}|/m~\ll~1$, neutron matter can be modeled as a uniform system 
of point-like particles of mass $m$,  the dynamics being dictated by the Hamiltonian (throughout this paper, we will
adopt a system of units such that $\hbar=c=1$)
\begin{equation}
\hat{H} = \sum_i -\frac{{\nabla}^2_i}{2m} + \sum_{j>i} \hat{v}_{ij} + \sum_{k>j>i} \hat{V}_{ijk}  \ .
\label{eq:hamiltonian}
\end{equation}
In the above equation, $\hat{v}_{ij}$ and $\hat{V}_{ijk}$ are the potentials describing two- and three-neutron interactions, 
while the ground and excited states appearing in Eq. \eqref{def:W}, $|\Psi_0\rangle$ and $|\Psi_n\rangle$, are eigenstates 
of $H$ belonging to the eigenvalues $E_0$ and $E_n$, respectively.

At leading order of the expansion in powers of $|\mathbf{q}|/m$, the vector and axial-vector components of the weak neutral currents 
reduce to the density and spin-density fluctuation operators, defined according to
\begin{align}
\label{w:op1}
J_{Z}^0 \to\, & \hat{O}_{\mathbf{q}}^\rho = \sum_i \hat{O}_{\mathbf{q}}^\rho(i)=\sum_i e^{i \mathbf{q} \cdot \mathbf{r}_i} \\
\label{w:op2}
{\bf J}_{Z} \to\,& \hat{O}_{\mathbf{q}}^{\bm \sigma} = \sum_i \hat{O}_{\mathbf{q}}^{\bm \sigma}(i)
=\sum_i e^{i \mathbf{q} \cdot \mathbf{r}_i} {\bm \sigma}_i\, .
\end{align}

Using the above expressions and choosing a coordinate system such that the $z$-axis is in the direction of $\mathbf{q}$, implying 
$q\equiv(\omega,0,0,|\mathbf{q}|)$, and $\mathbf{k}$ and $\mathbf{k}^\prime$ lie in the the  $xz$-plane, one can write the
scattering rate of Eq. (\ref{eq:scattering_rate}) in the simple form
\begin{align}
\label{eq:scattering_rate2}
W(\mathbf{q},\omega)& = \frac{G_{F}^2}{4\pi^2}  \Big\{ (1+\cos\theta) S^\rho(\mathbf{q},\omega) \\
\nonumber
& +C_A(1-\cos\theta) S^\sigma(\mathbf{q},\omega) + \frac{C_A}{2 E E^\prime} \left[ Q_{x}^2 S^{\sigma}_{xx}(\textbf{q},\omega) \right. \\
\nonumber
& \left. + (Q_{z}^2-|\mathbf{q}|^2)S^{\sigma}_{zz}(\textbf{q},\omega) + 2 Q_x Q_z S^{\sigma}_{xz}(\textbf{q},\omega)\right] \Big\} \ ,
\end{align}
where $\cos\theta=\hat{k}\cdot\hat{k}^\prime$, $Q = k+k^\prime\equiv(\Omega,Q_x,0,Q_z)$ and $C_A\simeq 1.25$ is the ratio of 
the weak axial vector and Fermi coupling constants of the nucleon. The density and spin-density response functions are defined by 
\begin{align}
\label{eq:dresp_def}
S^{\rho}(\textbf{q},\omega) &=\frac{1}{A} \sum_n |\langle \Psi_n | \hat{O}_{\textbf{q}}^{\rho}| \Psi_0\rangle |^2  \delta(\omega+E_0-E_n)  \ , \\
\label{eq:sresp_def}
S^{\sigma}_{\alpha\beta}(\textbf{q},\omega) &=\frac{1}{A} \sum_n \langle \Psi_n | \hat{O}^{\sigma_\alpha}_{\textbf{q}}| \Psi_0\rangle \langle \Psi_0 | \hat{O}^{\sigma_\beta}_{\textbf{q}} | \Psi_n\rangle   \\
\nonumber
&  \times  \delta(\omega+E_0-E_n) \  , 
\end{align}
$A$ being the particle number. 

Note that in Eq. \eqref{eq:scattering_rate2} we have introduced the trace of the spin-density response matrix
\beq
S^\sigma(\mathbf{q},\omega)=\sum_{\alpha}S^{\sigma}_{\alpha\alpha}(\mathbf{q},\omega) \ . 
\eeq
In the absence of non central
 interactions, $S^{\sigma}_{zz}=S^{\sigma}_{xx}=S^{\sigma}/3$, $S^{\sigma}_{xz}~=~0$,  and Eq. (\ref{eq:scattering_rate2}) 
reduces to  \cite{IW}
\begin{align}
\nonumber
W(\mathbf{q},\omega) =\frac{G_{F}^2}{4 \pi^2} & \left[ (1+\cos\theta)S^\rho(\mathbf{q},\omega) \right. \\
& \left. +\frac{C_{A}^2}{3}(3-\cos\theta) S^\sigma(\mathbf{q},\omega)\right] \, .
\end{align}

\subsection{Correlated Basis Functions}
\label{CBF}

Our work is based on the CBF formalism, in which the states appearing in Eq. (\ref{def:W}) are written
in the form \cite{CBF1,CBF2}
\begin{equation}
|\Psi_n\rangle \equiv\frac{\hat{\mathcal{F}}|\Phi_n\rangle}{\langle \Phi_n|\hat{\mathcal{F}}^\dagger \hat{\mathcal{F}} | \Phi_n\rangle},
\end{equation}
where $|\Phi_n\rangle$ is the Slater determinant describing a $n$ particle-$n$ hole state of the non interacting neutron gas. 

The operator $\hat{\mathcal{F}}$,  accounting for the correlation structure induced by 
NN interactions, is generally written in product form as
\begin{equation}
\mathcal{\hat{F}}=\mathcal{S} \prod_{j>i=1}^A \hat{F}_{ij} \ ,
\label{eq:Foperator}
\end{equation}
with the two-body correlation operator $\hat{F}_{ij}$ exhibiting a complex structure, which reflects the spin-dependence
and non central nature of nuclear forces in the neutron-neutron sector. As a consequence, $[\hat{F}_{ij},\hat{F}_{ik}] \neq 0$, and the product 
in the right hand side of Eq. \eqref{eq:Foperator} has to be properly symmetrized through the action of the operator $\mathcal{S}$.

The calculations discussed in this article have 
been carried out using the Argonne $v_{6}^\prime$ potential \cite{wiringa:02}, which can be written in the form
\beq
\hat{v}_{ij} = \sum_{p=1}^6 v^{p}(r_{ij})\hat{O}^{p}_{ij} \ ,
\label{v6prime}
\eeq
with
\begin{align}
\hat{O}^{p=1-6}_{ij}&=(1,\sigma_{ij},S_{ij})\otimes(1,\tau_{ij}) \ .
\end{align}
In the above equation, $\sigma_{ij}={\bm \sigma}_i \cdot {\bm \sigma}_j$ and $\tau_{ij}={\bm \tau}_i \cdot {\bm \tau}_j$, ${\bm \sigma}_i$ and ${\bm \tau}_i$ being Pauli matrices  
acting in the spin and isospin space of the $i$-th nucleon, respectively, while the tensor operator is 
\begin{equation}
S_{ij}=\frac{3}{{\bf r}_{ij}^2} ({\bf r}_{ij} \cdot {\bm \sigma}_i)  ({\bf r}_{ij} \cdot {\bm \sigma}_j) - ({\bm \sigma}_i \cdot {\bm \sigma}_j)  \ .
\label{eq:tens_def}
\end{equation}

The two-nucleon correlation operator is consistently defined as 
\begin{align}
\hat{F}_{ij}& = \sum_{p=1}^6 f^{p}(r_{ij})\hat{O}^{p}_{ij} \ ,
\label{eq:Foperator_v6prime}
\end{align}
and the shape of the correlation functions $f^{p}(r_{ij})$ is determined solving the Euler-Lagrange equations obtained from the minimization of 
the expectation value of the Hamiltonian in the CBF ground state 
\begin{equation}
\frac{\delta \langle \hat{H} \rangle}{\delta f^p}\equiv \frac{\delta}{ \delta f^p} \langle \Psi_ 0 | \hat{H} | \Psi_0 \rangle=0 \ .
\label{eq:min}
\end{equation}
  
The calculation of the Hamiltonian expectation value, and, more generally, of the matrix elements of a many-body operator between correlated states, 
involves severe difficulties. While very accurate calculations of $\langle \hat{H} \rangle$ can be carried out within stochastic approaches, such as
 Auxiliary Field Diffusion Monte Carlo (AFDMC) \cite{schmidt:99}, the derivation of the CBF effective interaction requires the use of the cluster 
expansion technique \cite{CBF1}, which allows one to cast the Hamiltonian expectation value in the form 
\beq
\label{clusterexp}
\langle \hat{H} \rangle = T_F + \sum_{n=2}^\infty  \langle \Delta \hat{H} \rangle_n \ , 
\eeq
where $T_F$ is the energy of the non interacting Fermi gas and $\langle \Delta \hat{H} \rangle_n$ denotes the contribution to $\langle \hat{H} \rangle$
arising from subsystems (clusters) involving $n$ interacting particles. The Euler Lagrange equations for the $f^{p}(r_{ij})$ are obtained at two-body 
cluster level, i.e. neglecting all contributions with $n\geq3$ in the right hand side of Eq. \eqref{clusterexp}. A detailed account of the procedure employed to  
determine the correlation functions can be found in Ref. \cite{Lovato13}.

\subsection{Effective interaction}
\label{veff}

The CBF effective interaction is defined by the relation
\begin{equation}
\langle  \hat{H}\rangle\equiv T_F+\langle \Phi_0 | \sum_{j>i} \hat{v}_{ij}^{\text{eff}} | \Phi_0 \rangle \ ,
\label{eq:eff_int}
\end{equation}
clearly showing that $v^{\rm eff}_{ij}$ depends on the level of approximation employed to evaluate  $\langle \hat{H}\rangle$ within 
the cluster expansion scheme. 

\begin{figure}[h!]
\begin{center}
\includegraphics[scale=0.70]{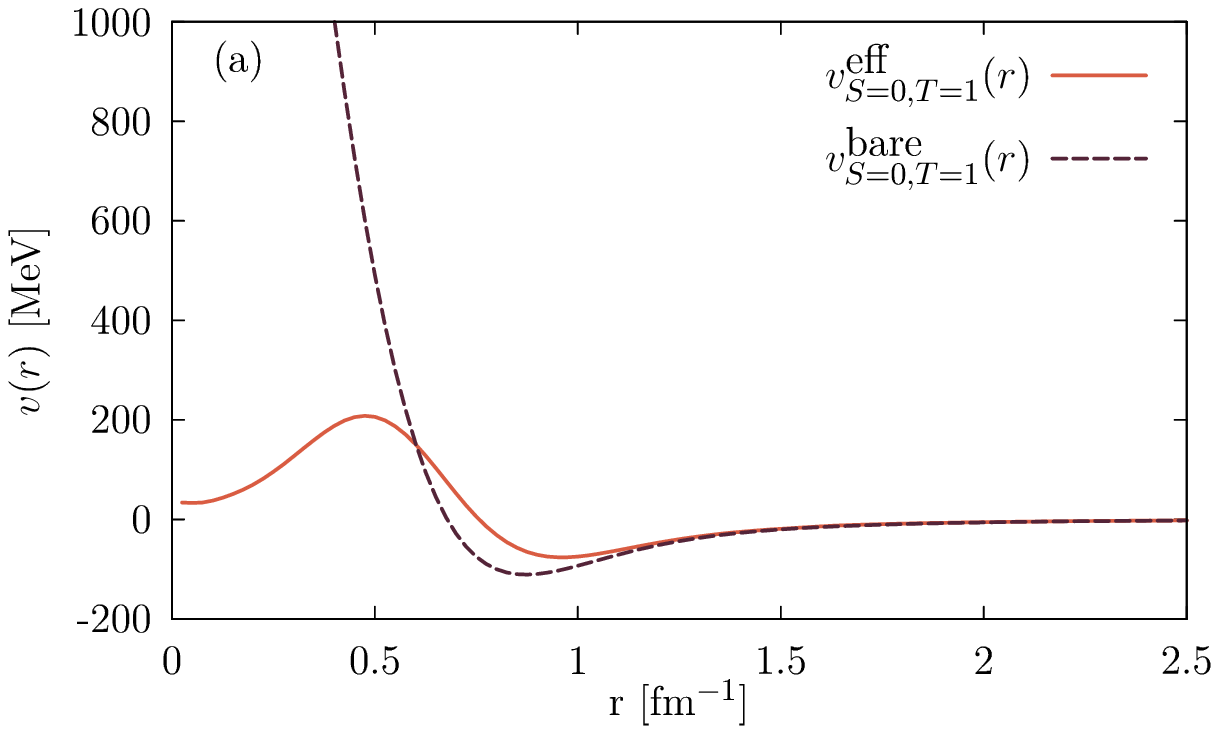}
\includegraphics[scale=0.70]{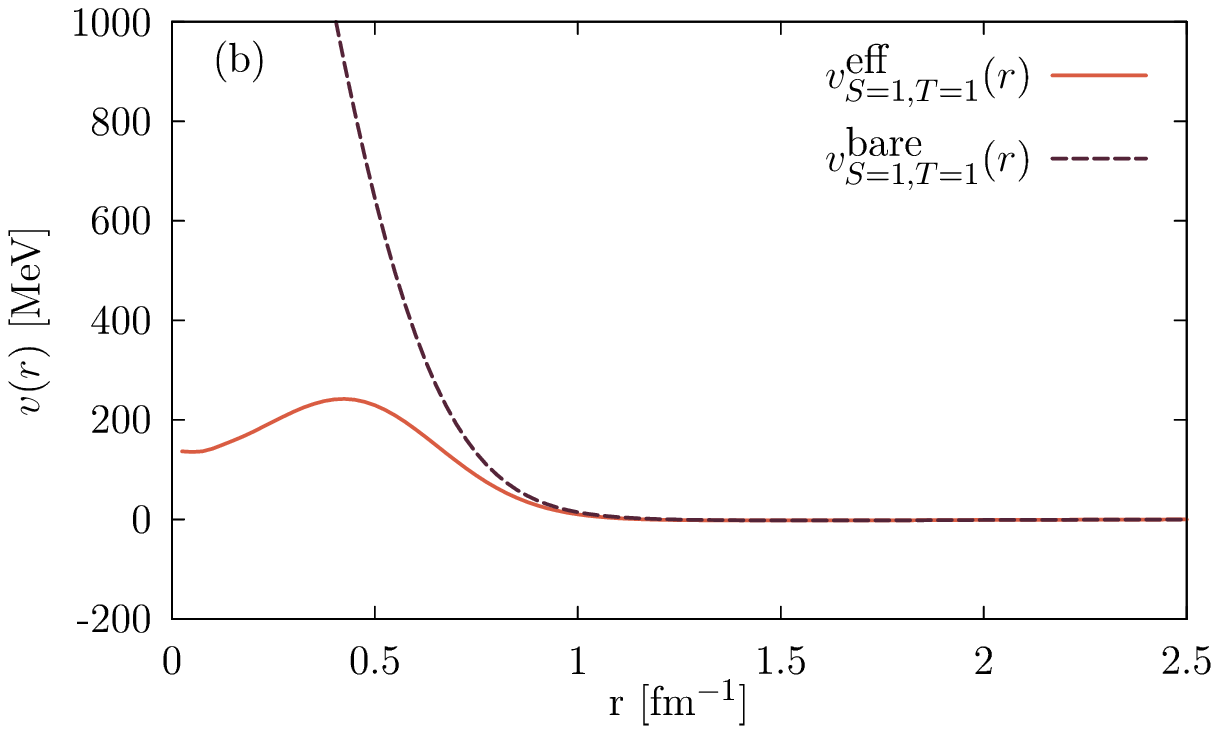}
\includegraphics[scale=0.70]{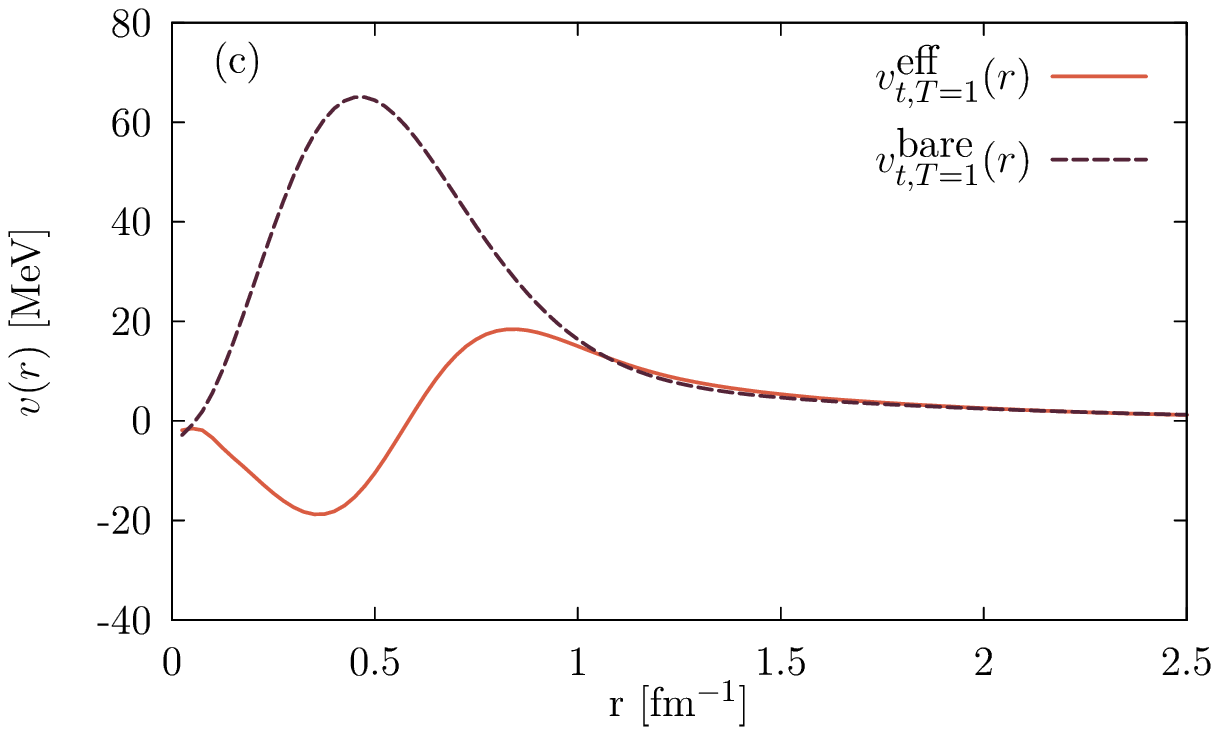}
\caption{(color online) (a) Comparison between the central component of the Argonne $v_{6}^\prime$ potential acting between a nucleon pair
coupled to total spin and isospin $S=0$ and $T=1$ (dashed line) and the corresponding CBF effective interaction, obtained 
in Ref. \cite{Lovato13} (solid line); (b) same as in (a) but for the $S=1$ and $T=1$ channel; (c) same as in (b), but for the tensor component.}
\label{fig:veff}
\end{center}
\end{figure}

The authors of Ref. \cite{shannon_veff} obtained the effective interaction including the two-nucleon cluster contribution only. This prescription, 
while leading to a very simple and transparent expression of $v^{\rm eff}_{ij}$, disregards the effects of three-nucleon forces, 
which are known to play a critical role in determining the energy spectrum of light nuclei, as well as the saturation properties of  
symmetric nuclear matter. In Refs. \cite{BV,Benhar_Farina}, interactions involving more than two nucleons have been taken into 
account within the approach originally proposed in Ref. \cite{LagPan}, in which the
main effect of three- and many-body forces is described through a density dependent modification
of the NN potential at intermediate range. 

A significantly improved CBF effective interaction, obtained including the three-nucleon cluster contribution in the calculation of 
the ground state expectation value of the Hamiltonian, has been developed
by the authors of Ref. \cite{Lovato13}. Within the scheme of Ref. \cite{Lovato13}, three-nucleon interactions are described 
at fully microscopic level, using the accurate parametrization of the potential referred to as UIX \cite{UIX}. 

In Fig. \ref{fig:veff}, the central and tensor components of the CBF effective interaction acting in the neutron-neutron channel 
are compared to the corresponding components of the Argonne $v_{6}^\prime$ potential. It clearly appears that the repulsive core  of the 
bare potential is wiped out by the screening effect arising from short range NN correlations. As a consequence,  the CBF effective interaction turns out to be  
well behaved, and can  be used to carry out perturbative calculations in the Fermi gas basis. Note that, 
because central and spin correlations are short ranged, the $S=0$ and $S=1$ central components of the effective interaction are
identical  to the corresponding components of the bare potential at $r \gtrsim 1.2$ fm. On the other hand, owing to the longer range of tensor correlations, 
the tail of the tensor component of the effective interaction lies slightly above the one of the Argonne $v_{6}^\prime$ potential. 

\begin{figure}[h!]
\vspace*{.2in}
\begin{center}
\includegraphics[scale=0.7]{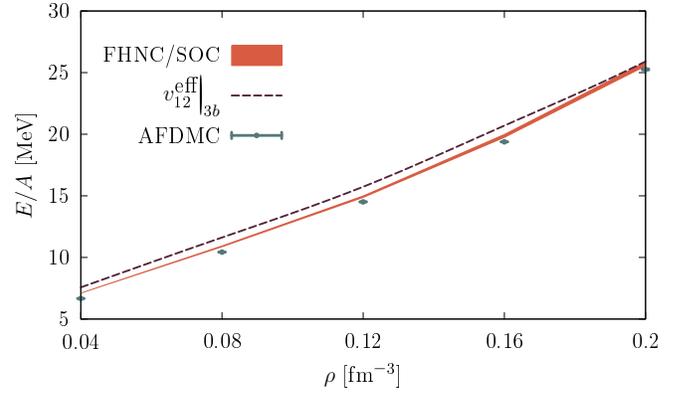}
\caption{(color online) Density dependence of the energy per particle of cold neutron matter. The dashed line shows the results obtained
using the CBF effective interaction of Ref. \cite{Lovato13} in the Hartree-Fock approximation, while the shaded region illustrates the  
uncertainty of the FHNC/SOC energies, arising from different treatments of the kinetic terms. Finally, the dots represent the AFDMC 
results obtained with the bare interactions.}
\label{fig:energy}
\end{center}
\end{figure}

The accuracy of the approach of Ref. \cite{Lovato13} is illustrated in Fig. \ref{fig:energy}, where the density-dependence of the energy per particle 
of pure neutron matter computed in  Hartree-Fock approximation using the CBF effective interaction, shown by the dashed line, is compared to 
the results obtained taking into account contributions at all orders  of the cluster expansion within the FHNC/SOC summation scheme \cite{FHNC/SOC}. The shaded region provides a measure of the uncertainty of the FHNC/SOC results associated with the treatment of the kinetic energy \cite{CBF1}. AFDMC energies, also shown in Fig. \ref{fig:energy}, are only slightly lower than the variational ones.  
\subsection{Effective operators}
\label{eff:op}

The effective operators are defined in terms of transition matrix elements through the equation [compare to Eq. \eqref{eq:eff_int}]
\begin{align}
\langle \Phi_n| \hat{O}_{\mathbf{q}}^{\text{eff}}| \Phi_0 \rangle \equiv \langle \Psi_n | \hat{O}_{\mathbf{q}} | \Psi_0 \rangle \ .
\label{eq:cbfOeff}
\end{align}
We have performed a calculation of the density and spin-density responses, Eqs. \eqref{eq:dresp_def} and \eqref{eq:sresp_def}, taking into account correlated one particle-one 
hole (1p-1h) final states only. In the 1p-1h sector, Eq. \eqref{eq:cbfOeff} reduces to
\begin{align}
&\langle \Phi_{p_m; h_i} | \hat{O}_{\mathbf{q}}^{\text{eff}}| \Phi_0 \rangle \equiv \langle \Psi_{p_m; h_i} | \hat{O}_{\mathbf{q}}  | \Psi_0\rangle \ ,
\label{eq:cbf_1p1heff}
\end{align}
where the labels $p_m$ and $h_i$ specify the quantum numbers of the particle and hole states involved in the transition, respectively. 

The matrix elements of the Fermi and Gamow-Teller 
operators between correlated states are obtained from a cluster expansion, similar to the one employed to evaluate the ground state expectation value of the Hamiltonian.
In this case, the smallness parameters are $f^{1}_{ij}-1$ and $f^{p>1}_{ij}$. 

\begin{figure}[h!]
\begin{center}
\includegraphics[scale=1.1]{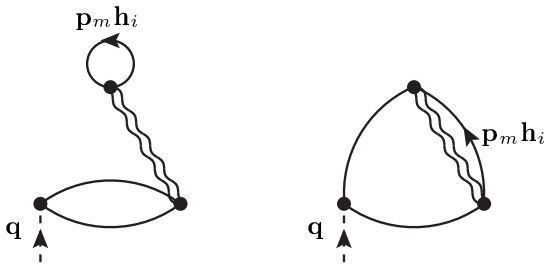}

\caption{Three body diagrams of second order in $f^{1}_{ij}-1$ and $f^{p>1}_{ij}$ emerging from the cluster expansion of the 1p-1h
transition matrix elements \cite{Lovato13}. Dynamical and statistical correlations are represented by wavy lines and oriented solid lines, 
respectively, while the oriented dashed line depicts the interaction with the external probe.  \label{fig:3b_2nd_order}}
\end{center}
\end{figure}

In Refs. \cite{shannon_2,Benhar_Farina} the effective operators have been obtained
including two-body cluster contributions only.   
In the present work, we follow the scheme of Ref.~\cite{Lovato13}, in which, consistently with the definition of the effective interaction,  three-body 
cluster contributions to the transition matrix elements are taken into account at leading order in $f^{1}_{ij}-1$ and $f^{p>1}_{ij}$. 
We have further improved on this procedure by adding the second order 
three-body diagrams shown in Fig. \ref{fig:3b_2nd_order}.

\section{Results}
\label{totresults}

\subsection{Density and spin-density responses}
\label{responses}

The density and spin-density responses have been computed from Eqs. \eqref{eq:dresp_def} and \eqref{eq:sresp_def}, respectively, within different approximation schemes.

In the Correlated Hartree-Fock (CHF) approximation the sum over final states is restricted to correlated 1p-1h states with excitation energy 
\beq
\omega_{i m}= e(p_m) - e(h_i) \ ,
\eeq 
where the single particle energies are computed using  the CBF effective interaction in Hartree-Fock approximation.

As pointed out in Ref. \cite{Benhar_Farina}, the inclusion of interaction effects within the CHF approximation takes into account both 
the appearance of a mean field, resulting in the departure of the single particle spectrum from the kinetic energy behavior, and the occurrence 
of short range NN correlations that move strength from the 1p-1h channel to more complex np-nh final states, thus 
leading to a sizable quenching of the transition amplitudes of Eq. \eqref{eq:cbf_1p1heff}.

While these effects are known to be dominant in the kinematical region corresponding to momentum transfer $|{\bf q}| \gsim 2 \ {\rm fm}^{-1}$ \cite{Benhar_Farina}, 
in which single nucleon knock out is the main reaction mechanism, at lower momentum transfer the occurrence of collective excitation also plays a critical role.
These excitation modes have been taken into account within the Correlated Tamm-Dancoff (CTD) approximation, which amounts 
to writing the final states in Eqs. \eqref{eq:dresp_def} and \eqref{eq:sresp_def} as a superposition of 1p-1h excitations of definite 
spin, $S$, and spin projection along the $z$-axis,  $S_z$, according to \cite{Lovato13}
\begin{align}
|\Phi_n\rangle_{S}^{CTD}=\sum_{\mathbf{p}_m\mathbf{h}_i S_z}C^{n\,SS_z}_{\mathbf{p}_m\mathbf{h}_i}
 |\Phi_{\mathbf{p}_m;\mathbf{h}_i}\rangle_{SS_z}\, .
 \label{CTD:states}
\end{align}
The coefficients appearing in the above expansion, as well as the spectrum of the excitation energies of the states, $\omega_n^S$, are determined solving the 
eigenvalue equations
\begin{align}
\hat{H}^{\text{eff}}|\Phi_n\rangle^{CTD}_S= (E_0+\omega_{n}^S)|\Phi_n\rangle^{CTD}_S\, ,
\label{eq:eigen_TDA}
\end{align}
with
\begin{align}
\hat{H}^{\text{eff}} = \sum_i-\frac{\nabla^{2}_i}{2m}+\sum_{j>i}\hat{v}^{\text{eff}}_{ij} \ .
\end{align}

Numerical calculations  have been performed on a cubic lattice, with a discrete set of $N_h$ states specified by the hole momenta
${\bf h}_i$ such that $|{\bf h}_i| < k_F$ and $|{\bf h}_i + {\bf q}| > k_F$, $k_F$ being the Fermi momentum.
The diagonalization of the Hamiltonian matrix 
has been carried out using $\sim$~30000 basis states, and the resulting responses have been converted to smooth functions of $\omega$ using
a Gaussian representation of the energy conserving $\delta$-function of finite width $\sigma$.

\begin{figure}[!h]
\begin{center}
\hspace*{-0.0cm}
\includegraphics[width=8.5cm,angle=0]{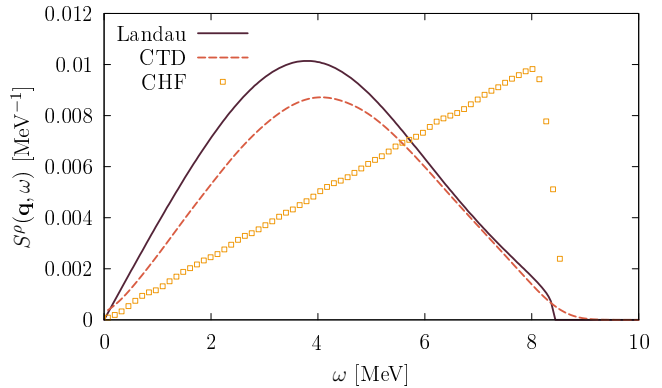}
\end{center}
\caption{
Density response function of pure  neutron matter at $\rho = \rho_0 = 0.16 \ {\rm fm}^{-3}$ for momentum transfer $q=0.1\ {\rm fm}^{-1}$. The dashed line and the squares show the results of the CTD and CHF approximations, respectively, while the solid line has been obtained within the framework of Landau theory of normal Fermi liquids \cite{BCL}. \label{fig:landau_density}}
\end{figure}

The CTD density response of pure neutron matter at density $\rho = \rho_0$, $\rho_0 = 0.16 \ {\rm fm}^{-3}$ being the equilibrium density 
of isospin symmetric matter, is shown by the dashed line of Fig. \ref{fig:landau_density}. For comparison, we also show, by the solid line, 
the results obtained from Landau theory of normal Fermi liquids using values of the Landau parameters extracted from the matrix elements
of the CBF effective interaction \cite{BCL}. 

It appears that, unlike the case of isospin symmetric matter discussed in Ref. \cite{Lovato13}, in pure neutron matter the density response does not exhibit the sharp peak 
arising from the excitation of the collective mode. We also note that two conceptually different approaches yield rather similar results, although the maximum of the CTD response turns 
out to be lower by $\sim$ 20 \%.
\begin{figure*}
\begin{center}
\hspace*{-0.4cm}
\includegraphics[width=11.0cm,angle=0]{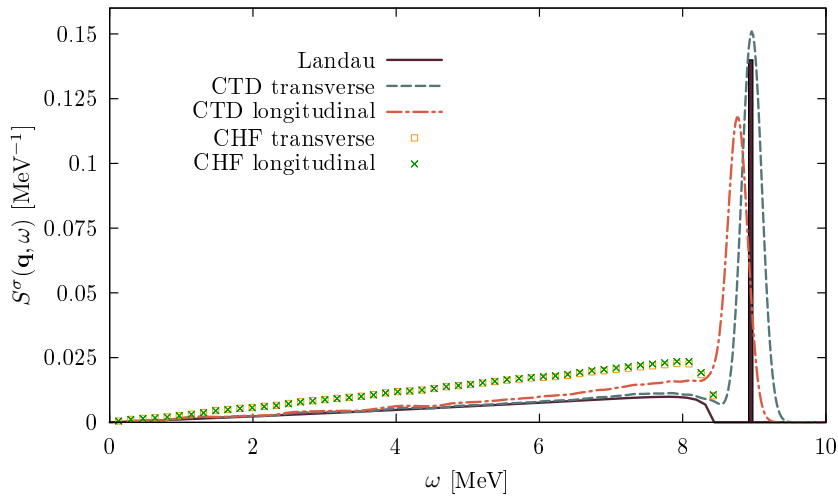}
\end{center}
\caption{ Spin-transverse (dashed line) and spin-longitudinal (dot-dash line) responses of pure neutron matter, computed within the CTD and CHF approximations 
at $\rho = \rho_0 = 0.16 \ {\rm fm}^{-3}$ for momentum transfer $q=0.1\ {\rm fm}^{-1}$. The solid line has been obtained from Landau theory, according to the approach of Refs. \cite{IW,BCL}
 \label{fig:landau_spin}.}
\end{figure*}

In order to identify the contribution of tensor forces, the spin-density response function \eqref{eq:sresp_def} can be split into its longitudinal and transverse components, defined as 
\begin{align}
S_L(q,\omega)&=\frac{1}{A} \sum_n |\langle \Psi_n | \hat{\bf q} \cdot \hat{O}_{\textbf{q}}^{\bm \sigma}| \Psi_0\rangle |^2 \delta(\omega+E_0-E_n)  \ , \\
S_T(q,\omega)&=\frac{1}{A} \sum_n |\langle \Psi_n | \hat{\bf q} \times  \hat{O}_{\textbf{q}}^{\bm \sigma}| \Psi_0\rangle |^2 \delta(\omega+E_0-E_n) \ ,
\end{align}
where $\hat{\bf q} = {\bf q}/ |{\bf q}|$. Note that in the absence of tensor forces  $S_L=S_{zz}=S_{xx}=S_{T}$, 
the elements of the spin-density response matrix being defined as in Section \ref{NC}.

The spin-transverse and spin-longitudinal response functions of pure neutron matter at density $\rho = \rho_0 = 0.16 \ {\rm fm}^{-3}$, computed in CTD approximation, are displayed 
by the dashed and dot-dash lines of Fig.~\ref{fig:landau_spin}, respectively. The solid line corresponds to the spin-density response obtained from Landau theory within the 
approach discussed in Refs. \cite{IW,BCL}, in which the effect of the tensor interaction is neglected. 

In this case the peak associated with the collective excitation sticks out in both the spin-longitudinal and spin-transverse channels. The results of Landau theory are in perfect agreement 
with the CTD $S_T$, while tensor interactions turn out to have an appreciable effect on $S_L$.

\subsection{Neutrino mean free path}
\label{mfp}

\begin{figure}[h!]
\begin{center}
\hspace*{-0.4cm}
\includegraphics[width=8.2cm,angle=0]{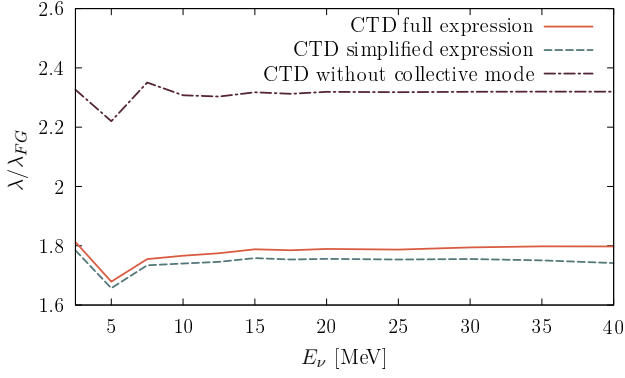}
\end{center}
\caption{ Energy dependence of the 
ratio  between the neutrino mean free path in pure neutron matter obtained within the approximation schemes discussed in Section \ref{responses} and that  corresponding 
to the non interacting Fermi gas.  Solid line: full CTD approximation; dashed line: CTD approximation without tensor interactions; dot-dash line: CHF approximation.
All calculations have been carried out at density $\rho = \rho_0$. \label{MFP}}
\end{figure}

The mean free path of non degenerate neutrinos, $\lambda$,  can be obtained from the scattering rate of Eq. \eqref{eq:scattering_rate2} using the relation \cite{IW}
\begin{align}
\label{def:lambda}
\lambda^{-1} = \rho \int \frac{d^3 q}{(2 \pi)^3} \ W({\bf q},\omega) \ .
\end{align}
The ratio between the mean free path at $\rho = \rho_0$ obtained within the approximation discussed in Section \ref{responses} and that corresponding to the non interacting 
Fermi gas, $\lambda_{FG}$,  is displayed in Fig.~\ref{MFP} as a function of neutrino energy. The solid line shows the results of the full CTD calculation, whereas
the dashed line has been obtained neglecting the effects of tensor forces. The dot-dash line represents the mean free path computed within the CHF approximations, 
i.e. not taking into account the occurrence of collective excitations. It clearly appears that interactions lead to a large enhancement of the mean free path, reflecting 
the quenching of the scattering rate, and that both short and long range correlations play an important role. In particular, the inclusion of the collective mode in the calculation
of the response produces a reduction of the neutrino mean free path of about $25\%$.

\subsection{Sum rules}
\label{sumrules}

As pointed out above, in the presence of NN correlations the set of final states accessible in neutrino nucleus interactions driven by the 
the operators of Eqs.~\eqref{w:op1} and \eqref{w:op2} is not exhausted by 1p-1h states. More complex np-nh states, not taken into 
account in this work, can also be excited.

The uncertainty associated with the truncation of
the space of final states can be estimated studying
the static structure functions, or sum rules, defined as the $\omega$-integrals of the corresponding response 
functions. 

Exploiting completeness of the set of final states, the density and spin-density sum rules can be cast in the form 
\begin{align}
\label{SR:D}
S^\rho(\mathbf{q})&= \int d \omega  \ S^\rho(\mathbf{q},\omega) \\
& = 1+ \rho \int d\mathbf{r}_{12} e^{i \mathbf{q}\cdot \mathbf{r}_{12}} [g^{c}(r_{12})-1]\nonumber\\
\label{SR:S}
S^\sigma(\mathbf{q})&=  \frac{1}{3}  \int d \omega  \ S^\sigma(\mathbf{q},\omega) \\ 
\nonumber
& = 1+\frac{1}{3} \rho\int d\mathbf{r}_{12} e^{i \mathbf{q}\cdot \mathbf{r}_{12}} g^{\sigma}(r_{12})\, ,
\end{align}
where the central and spin distribution functions are defined as \cite{SGRC}
\begin{align}
\label{gc}
g^c(r) & = \frac{1}{A} \ \frac{1}{2 \pi r^2 \rho} \ \sum_{j<i} \   \langle  \ \delta(r_{ij}-r)  \  \rangle  \ , \\
\label{gsig}
g^\sigma(r) & = \frac{1}{A} \ \frac{1}{2 \pi r^2 \rho} \ \sum_{j<i} \  \langle \  \delta(r_{ij}-r) ({\bm \sigma}_i \cdot {\bm \sigma}_j) \  \rangle \ ,
\end{align}
and $\langle \ \ldots \ \rangle$ denotes the expectation value in the neutron matter ground state. As the distribution functions \eqref{gc} and \eqref{gsig} can be
accurately evaluated within microscopic many body approaches using the same Hamiltonian employed in the CTD calculation, comparison between 
the integrated CTD responses, given by the first line of Eqs.~\eqref{SR:D} and \eqref{SR:S}, and the sum rules obtained from the second line  
provides a quantitative  estimate of the role of np-nh final states.   

\begin{figure}[h!]
\begin{center}
\hspace*{-0.4cm}
\includegraphics[width=8.2cm,angle=0]{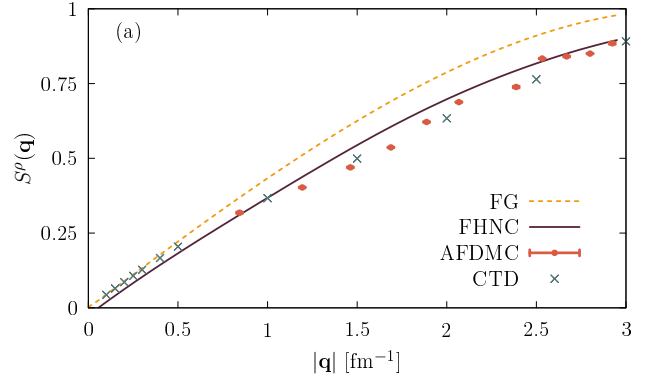}
\end{center}
\caption{ Density sum rule of pure neutron matter at $\rho = \rho_0$, as a function of the magnitude of the momentum transfer. The crosses show the results of the direct  
integration of the CTD response, whereas  the dashed line and the dots have been obtained computing the ground state expectation value  
of Eq. \eqref{gc} within the variational FHNC and AFDMC approaches, respectively. For comparison, the density 
sum rule of the non interacting Fermi gas is also shown by the solid line. \label{SRD}}
\end{figure}

\begin{figure}[h!]
\begin{center}
\hspace*{-0.4cm}
\includegraphics[width=8.2cm,angle=0]{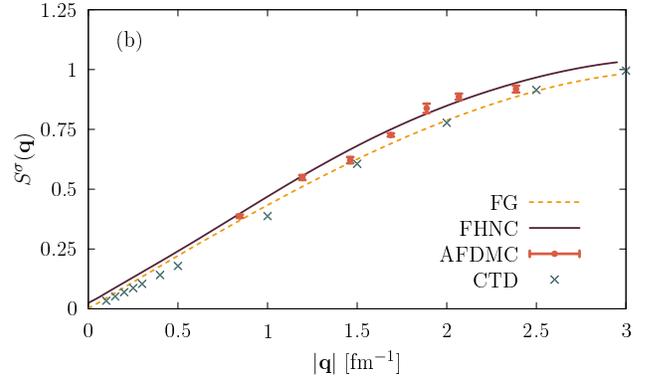}
\end{center}
\caption{Same as in Fig. \ref{SRD}, but for the spin-density sum rule.  The dashed line and the dots signs have been obtained computing the ground state expectation value  
of Eq. \eqref{gsig}. \label{SRS}}
\end{figure}

In Figs. \ref{SRD} and \ref{SRS} the sum rules extracted from the CTD responses (crosses) are compared to those computed using 
the FHNC distribution functions (solid lines).  We also show  the results of the 
AFDMC approach (dots with error bars), in which $S^\rho(\mathbf{q})$ and $S^\sigma(\mathbf{q})$ are 
obtained from a direct calculation of  the ground state expectation value of the operators~\eqref{w:op1} and~\eqref{w:op2}. 

A detailed description of the evaluation of observables within AFDMC can be found in, e.g., 
Refs.~\cite{gandolfi:09,gandolfi:09b}.
Because in the case of operators that do not commute with the Hamiltonian
the result depends on the trial wave function, we have computed the spin-density sum rule
exploiting the Helmann-Feynmann theorem and using the modified Hamiltonian
\begin{equation}
\hat H^\prime_\alpha =\hat H+2\,\alpha\sum_{j>i}\cos({\mathbf{q}\cdot \bm r}_{ij})\,{\bm\sigma}_i\cdot{\bm\sigma}_j \ .
\end{equation}
For each value of ${\bm q}$ the simulation has been performed several times using small values 
of $\alpha$, and the sum rule has been obtained from
\begin{equation}
S^\sigma(\mathbf{q}) - 1= \lim_{\alpha\rightarrow0}\frac{\partial E_{\alpha}}{\partial\alpha} \ , 
\end{equation}
where $E_{\alpha} = \langle \hat H^\prime_\alpha \rangle$.

The CTD sum rule of the density response exhibits the correct behaviour in the $|{\bf q}| \to 0$ 
limit, while, owing to the approximations involved in the variational approach, the FHNC 
result fails to fulfill the requirement $S^\rho(0) = 0$. It is worth mentioning that such violation
is in fact small: $S^\rho(0) \simeq 0.005$.
Due to the presence of tensor interactions, the spin-density sum rule is not constrained to vanish for vanishing momentum transfer. 
The FHNC calculation yields $S^\sigma(0) = 0.03$, while the CTD results appears to tend to zero.
The agreement between the CTD and AFDMC sum rules at intermediate momentum 
transfer ($|{\bf q}| \lsim 1.5 \ {\rm fm}^{-1}$)
suggests that in this kinematical region, in which the non relativistic approximation 
employed in our work is fully justified,  the contribution of 2p-2h final states is not large.


\section{Conclusions}
\label{conclusions}

We have carried out a calculation of the density and spin-density responses of pure neutron matter, 
determining the neutral current neutrino cross section in the low energy limit.

Our study is based on the CBF formalism, allowing for a fully consistent derivation of both the 
effective interaction and the effective operators within a dynamical model based on a realistic 
phenomenological Hamiltonian.

The role played by short and long range correlations has been analyzed comparing the results
of the CHF approximation, in which the space of neutron matter final states is
truncated at the level of 1p-1h excitations,  to those obtained within the CTD scheme, in which 
the transitions between 1p-1h states induced by the effective interaction are also taken into account.

Compared to the results of Ref. \cite{Lovato13}, in which a  similar study was carried out for 
isospin symmetric matter, we find an important qualitative difference, in that the excitation of the 
collective mode only occurs in the CTD spin-density response. 

To test the consistency between different many-body approaches based on the same dynamical model
we have compared the  CTD responses to the corresponding results obtained from Landau theory
using the set of Landau parameters computed using the CBF effective interaction. Overall, 
the agreement between the two conceptually different schemes turns out to be remarkably good.

The calculated responses have been used to determine the neutrino mean free path, 
providing a useful parametrization of  the opacity of neutron matter. The results clearly 
show that neutron--neutron interactions lead to a large increase of the mean free path, and that 
collective excitations play an important role.

The comparison of the $\omega$-integrated response functions with the sum rules   
obtained from accurate many-body calculations, carried out within the FHNC and AFDMC 
approaches, suggests that the contribution of 2p-2h final states, not included in our study, 
while being not totally negligible, is quite small in the kinematical region relevant to 
astrophysical applications, in which the non relativistic reduction of the nucleon weak 
current is expected to be applicable. However, the extension of the formalism described in this article 
to include the occurrence of 2p-2h final state does not involve severe 
conceptual difficulties.  

\section{Acknowledgements}
We thank R.B. Wiringa for carefully reading our manuscript.
This research is supported by the U.S.~Department of Energy, Office of
Nuclear Physics, under contracts DE-AC02-06CH11357 (A.L.),
DE-AC02-05CH11231 (S.G.) and by the NUCLEI SciDAC program. The work of
S.G. is also supported by the LANL LDRD program.
The computing time has been provided by Los Alamos Open Supercomputing. 
This research used also resources of the National Energy Research Scientific Computing
of the U.S. Department of Energy under Contract No.DE-AC02-05CH11231.
The work of O.B. is supported by INFN under grants MB31 and OG51.

\bibliographystyle{apsrev}

\bibliography{biblio}

\begin{thebibliography}{19}
\expandafter\ifx\csname natexlab\endcsname\relax\def\natexlab#1{#1}\fi
\expandafter\ifx\csname bibnamefont\endcsname\relax
  \def\bibnamefont#1{#1}\fi
\expandafter\ifx\csname bibfnamefont\endcsname\relax
  \def\bibfnamefont#1{#1}\fi
\expandafter\ifx\csname citenamefont\endcsname\relax
  \def\citenamefont#1{#1}\fi
\expandafter\ifx\csname url\endcsname\relax
  \def\url#1{\texttt{#1}}\fi
\expandafter\ifx\csname urlprefix\endcsname\relax\def\urlprefix{URL }\fi
\providecommand{\bibinfo}[2]{#2}
\providecommand{\eprint}[2][]{\url{#2}}

\bibitem[{\citenamefont{Cowell and Pandharipande}(2004)}]{shannon_veff}
\bibinfo{author}{\bibfnamefont{S.}~\bibnamefont{Cowell}} \bibnamefont{and}
  \bibinfo{author}{\bibfnamefont{V.~R.} \bibnamefont{Pandharipande}},
  \bibinfo{journal}{Phys. Rev. C} \textbf{\bibinfo{volume}{70}},
  \bibinfo{pages}{035801} (\bibinfo{year}{2004}).

\bibitem[{\citenamefont{Benhar and Valli}(2007)}]{BV}
\bibinfo{author}{\bibfnamefont{O.}~\bibnamefont{Benhar}} \bibnamefont{and}
  \bibinfo{author}{\bibfnamefont{M.}~\bibnamefont{Valli}},
  \bibinfo{journal}{Phys. Rev. Lett.} \textbf{\bibinfo{volume}{99}},
  \bibinfo{pages}{232501} (\bibinfo{year}{2007}),
  \urlprefix\url{http://link.aps.org/doi/10.1103/PhysRevLett.99.232501}.

\bibitem[{\citenamefont{{Lovato} et~al.}(2013)\citenamefont{{Lovato}, {Losa},
  and {Benhar}}}]{Lovato13}
\bibinfo{author}{\bibfnamefont{A.}~\bibnamefont{{Lovato}}},
  \bibinfo{author}{\bibfnamefont{C.}~\bibnamefont{{Losa}}}, \bibnamefont{and}
  \bibinfo{author}{\bibfnamefont{O.}~\bibnamefont{{Benhar}}},
  \bibinfo{journal}{Nuclear Physics A} \textbf{\bibinfo{volume}{901}},
  \bibinfo{pages}{22} (\bibinfo{year}{2013}), \eprint{1210.2099}.

\bibitem[{\citenamefont{Benhar et~al.}(2010)\citenamefont{Benhar, Polls, Valli,
  and Vida\~na}}]{Gmat}
\bibinfo{author}{\bibfnamefont{O.}~\bibnamefont{Benhar}},
  \bibinfo{author}{\bibfnamefont{A.}~\bibnamefont{Polls}},
  \bibinfo{author}{\bibfnamefont{M.}~\bibnamefont{Valli}}, \bibnamefont{and}
  \bibinfo{author}{\bibfnamefont{I.}~\bibnamefont{Vida\~na}},
  \bibinfo{journal}{Phys. Rev. C} \textbf{\bibinfo{volume}{81}},
  \bibinfo{pages}{024305} (\bibinfo{year}{2010}),
  \urlprefix\url{http://link.aps.org/doi/10.1103/PhysRevC.81.024305}.

\bibitem[{\citenamefont{{Benhar} and {Farina}}(2009)}]{Benhar_Farina}
\bibinfo{author}{\bibfnamefont{O.}~\bibnamefont{{Benhar}}} \bibnamefont{and}
  \bibinfo{author}{\bibfnamefont{N.}~\bibnamefont{{Farina}}},
  \bibinfo{journal}{Physics Letters B} \textbf{\bibinfo{volume}{680}},
  \bibinfo{pages}{305} (\bibinfo{year}{2009}), \eprint{0904.2260}.

\bibitem[{\citenamefont{Benhar et~al.}(2013)\citenamefont{Benhar, Cipollone,
  and Loreti}}]{BCL}
\bibinfo{author}{\bibfnamefont{O.}~\bibnamefont{Benhar}},
  \bibinfo{author}{\bibfnamefont{A.}~\bibnamefont{Cipollone}},
  \bibnamefont{and} \bibinfo{author}{\bibfnamefont{A.}~\bibnamefont{Loreti}},
  \bibinfo{journal}{Phys. Rev. C} \textbf{\bibinfo{volume}{87}},
  \bibinfo{pages}{014601} (\bibinfo{year}{2013}),
  \urlprefix\url{http://link.aps.org/doi/10.1103/PhysRevC.87.014601}.

\bibitem[{\citenamefont{Cipollone}(2012)}]{andrea_thesis}
\bibinfo{author}{\bibfnamefont{A.}~\bibnamefont{Cipollone}},
  \emph{\bibinfo{title}{Neutrino Interactions in Neutron Matter}}
  (\bibinfo{year}{2012}), \bibinfo{note}{phD Thesis, ``Sapienza'' Universit\`a
  di Roma}.

\bibitem[{\citenamefont{Iwamoto and Pethick}(1982)}]{IW}
\bibinfo{author}{\bibfnamefont{N.}~\bibnamefont{Iwamoto}} \bibnamefont{and}
  \bibinfo{author}{\bibfnamefont{C.~J.} \bibnamefont{Pethick}},
  \bibinfo{journal}{Phys. Rev. D} \textbf{\bibinfo{volume}{25}},
  \bibinfo{pages}{313} (\bibinfo{year}{1982}),
  \urlprefix\url{http://link.aps.org/doi/10.1103/PhysRevD.25.313}.

\bibitem[{\citenamefont{Clark}(1979)}]{CBF1}
\bibinfo{author}{\bibfnamefont{J.~W.} \bibnamefont{Clark}},
  \bibinfo{journal}{Progress in Particle and Nuclear Physics}
  \textbf{\bibinfo{volume}{2}}, \bibinfo{pages}{89 } (\bibinfo{year}{1979}),
  ISSN \bibinfo{issn}{0146-6410},
  \urlprefix\url{http://www.sciencedirect.com/science/article/pii/0146641079900048}.

\bibitem[{\citenamefont{Fantoni and Pandharipande}(1988)}]{CBF2}
\bibinfo{author}{\bibfnamefont{S.}~\bibnamefont{Fantoni}} \bibnamefont{and}
  \bibinfo{author}{\bibfnamefont{V.~R.} \bibnamefont{Pandharipande}},
  \bibinfo{journal}{Phys. Rev. C} \textbf{\bibinfo{volume}{37}},
  \bibinfo{pages}{1697} (\bibinfo{year}{1988}),
  \urlprefix\url{http://link.aps.org/doi/10.1103/PhysRevC.37.1697}.

\bibitem[{\citenamefont{Wiringa and Pieper}(2002)}]{wiringa:02}
\bibinfo{author}{\bibfnamefont{R.~B.} \bibnamefont{Wiringa}} \bibnamefont{and}
  \bibinfo{author}{\bibfnamefont{S.~C.} \bibnamefont{Pieper}},
  \bibinfo{journal}{Phys. Rev. Lett.} \textbf{\bibinfo{volume}{89}},
  \bibinfo{pages}{182501} (\bibinfo{year}{2002}),
  \urlprefix\url{http://link.aps.org/doi/10.1103/PhysRevLett.89.182501}.

\bibitem[{\citenamefont{Schmidt and Fantoni}(1999)}]{schmidt:99}
\bibinfo{author}{\bibfnamefont{K.}~\bibnamefont{Schmidt}} \bibnamefont{and}
  \bibinfo{author}{\bibfnamefont{S.}~\bibnamefont{Fantoni}},
  \bibinfo{journal}{Physics Letters B} \textbf{\bibinfo{volume}{446}},
  \bibinfo{pages}{99 } (\bibinfo{year}{1999}), ISSN \bibinfo{issn}{0370-2693},
  \urlprefix\url{http://www.sciencedirect.com/science/article/pii/S0370269398015226}.

\bibitem[{\citenamefont{Lagaris and Pandharipande}(1981)}]{LagPan}
\bibinfo{author}{\bibfnamefont{I.}~\bibnamefont{Lagaris}} \bibnamefont{and}
  \bibinfo{author}{\bibfnamefont{V.}~\bibnamefont{Pandharipande}},
  \bibinfo{journal}{Nuclear Physics A} \textbf{\bibinfo{volume}{359}},
  \bibinfo{pages}{349 } (\bibinfo{year}{1981}), ISSN \bibinfo{issn}{0375-9474},
  \urlprefix\url{http://www.sciencedirect.com/science/article/pii/0375947481902414}.

\bibitem[{\citenamefont{Pudliner et~al.}(1995)\citenamefont{Pudliner,
  Pandharipande, Carlson, and Wiringa}}]{UIX}
\bibinfo{author}{\bibfnamefont{B.~S.} \bibnamefont{Pudliner}},
  \bibinfo{author}{\bibfnamefont{V.~R.} \bibnamefont{Pandharipande}},
  \bibinfo{author}{\bibfnamefont{J.}~\bibnamefont{Carlson}}, \bibnamefont{and}
  \bibinfo{author}{\bibfnamefont{R.~B.} \bibnamefont{Wiringa}},
  \bibinfo{journal}{Phys. Rev. Lett.} \textbf{\bibinfo{volume}{74}},
  \bibinfo{pages}{4396} (\bibinfo{year}{1995}),
  \urlprefix\url{http://link.aps.org/doi/10.1103/PhysRevLett.74.4396}.

\bibitem[{\citenamefont{Pandharipande and Wiringa}(1979)}]{FHNC/SOC}
\bibinfo{author}{\bibfnamefont{V.~R.} \bibnamefont{Pandharipande}}
  \bibnamefont{and} \bibinfo{author}{\bibfnamefont{R.~B.}
  \bibnamefont{Wiringa}}, \bibinfo{journal}{Rev. Mod. Phys.}
  \textbf{\bibinfo{volume}{51}}, \bibinfo{pages}{821} (\bibinfo{year}{1979}),
  \urlprefix\url{http://link.aps.org/doi/10.1103/RevModPhys.51.821}.

\bibitem[{\citenamefont{Cowell and Pandharipande}(2003)}]{shannon_2}
\bibinfo{author}{\bibfnamefont{S.}~\bibnamefont{Cowell}} \bibnamefont{and}
  \bibinfo{author}{\bibfnamefont{V.~R.} \bibnamefont{Pandharipande}},
  \bibinfo{journal}{Phys. Rev. C} \textbf{\bibinfo{volume}{67}},
  \bibinfo{pages}{035504} (\bibinfo{year}{2003}),
  \urlprefix\url{http://link.aps.org/doi/10.1103/PhysRevC.67.035504}.

\bibitem[{\citenamefont{Shen et~al.}(2013)\citenamefont{Shen, Gandolfi, Reddy,
  and Carlson}}]{SGRC}
\bibinfo{author}{\bibfnamefont{G.}~\bibnamefont{Shen}},
  \bibinfo{author}{\bibfnamefont{S.}~\bibnamefont{Gandolfi}},
  \bibinfo{author}{\bibfnamefont{S.}~\bibnamefont{Reddy}}, \bibnamefont{and}
  \bibinfo{author}{\bibfnamefont{J.}~\bibnamefont{Carlson}},
  \bibinfo{journal}{Phys. Rev. C} \textbf{\bibinfo{volume}{87}},
  \bibinfo{pages}{025802} (\bibinfo{year}{2013}),
  \urlprefix\url{http://link.aps.org/doi/10.1103/PhysRevC.87.025802}.

\bibitem[{\citenamefont{Gandolfi
  et~al.}(2009{\natexlab{a}})\citenamefont{Gandolfi, Illarionov, Schmidt,
  Pederiva, and Fantoni}}]{gandolfi:09}
\bibinfo{author}{\bibfnamefont{S.}~\bibnamefont{Gandolfi}},
  \bibinfo{author}{\bibfnamefont{A.~Y.} \bibnamefont{Illarionov}},
  \bibinfo{author}{\bibfnamefont{K.~E.} \bibnamefont{Schmidt}},
  \bibinfo{author}{\bibfnamefont{F.}~\bibnamefont{Pederiva}}, \bibnamefont{and}
  \bibinfo{author}{\bibfnamefont{S.}~\bibnamefont{Fantoni}},
  \bibinfo{journal}{Phys. Rev. C} \textbf{\bibinfo{volume}{79}},
  \bibinfo{pages}{054005} (\bibinfo{year}{2009}{\natexlab{a}}),
  \urlprefix\url{http://link.aps.org/doi/10.1103/PhysRevC.79.054005}.

\bibitem[{\citenamefont{Gandolfi
  et~al.}(2009{\natexlab{b}})\citenamefont{Gandolfi, Illarionov, Pederiva,
  Schmidt, and Fantoni}}]{gandolfi:09b}
\bibinfo{author}{\bibfnamefont{S.}~\bibnamefont{Gandolfi}},
  \bibinfo{author}{\bibfnamefont{A.~Y.} \bibnamefont{Illarionov}},
  \bibinfo{author}{\bibfnamefont{F.}~\bibnamefont{Pederiva}},
  \bibinfo{author}{\bibfnamefont{K.~E.} \bibnamefont{Schmidt}},
  \bibnamefont{and} \bibinfo{author}{\bibfnamefont{S.}~\bibnamefont{Fantoni}},
  \bibinfo{journal}{Phys. Rev. C} \textbf{\bibinfo{volume}{80}},
  \bibinfo{pages}{045802} (\bibinfo{year}{2009}{\natexlab{b}}),
  \urlprefix\url{http://link.aps.org/doi/10.1103/PhysRevC.80.045802}.

\end{thebibliography}

\end{document}